\documentclass[sigconf]{acmart}
\AtBeginDocument{%
  }

\usepackage{enumitem} 
\usepackage{graphicx}
\usepackage{subcaption}

\acmConference[WWW '27]{Proceedings of the ACM Web Conference 2027}{May 10--14, 2027}{Dublin, Ireland}
\begin{document}

% \title{GR4AD-V2: Scaling Generative Advertising Recommendation from Semantic Representation to Value-Aligned Generation}
% \title{GR4AD-V2: Scaling Generative Advertising Recommendation by Bridging Semantic Representation and Commercial Value}
% \title{GR4AD-V2: Reliable, Scalable, and Value-Aligned Generative Recommendation for Large-Scale Advertising}
% \title{GR4AD-V2: Production-Native Generative Recommendation for Large-Scale Advertising}
\title{Towards Reliable, Scalable, and Value-Aligned Generative Recommendation for Large-Scale Advertising}

% Compact author block: two centered rows, with explicit contribution marks.
% Email addresses are placeholders; replace them before submission.
\author{%
  Yatong Xiao\textsuperscript{*},
  Yibo Sheng\textsuperscript{*},
  Lixiang Wang\textsuperscript{*},
  Mingjie Sun\textsuperscript{*},
  Yunhao Sha\textsuperscript{*}\\
  Pengfei Zhang\textsuperscript{*},
  Shengyu Wang\textsuperscript{*},
  Peng Wang\textsuperscript{*},
  Wenjin Wu,
  Peng Jiang\texorpdfstring{\textsuperscript{\ensuremath{\dagger}}}{}%
}
\affiliation{%
  \institution{Kuaishou Technology}
  \city{Beijing}
  \country{China}
  \\{
    \{xiaoyatong, shengyibo, wanglixiang03, sunmingjie, shayunhao,}%
  \\{
    zhangpengfei05, wangshengyu, wangpeng16, wuwenjin, jiangpeng\}@kuaishou.com}%
}
\renewcommand{\shortauthors}{Xiao et al.}

\begin{abstract}

Generative recommendation (GR) reformulates item retrieval as autoregressive generation of discrete semantic identifiers (SIDs), offering a unified paradigm for personalized recommendation. GR4AD, as an industrial generative recommendation framework, has greatly improved advertising revenue while enabling efficient real-time serving in Kuaishou. However, continued deployment has exposed three limitations: deterministic SID assignment remains sensitive to semantic boundaries; conventional backbone scaling provides diminishing revenue gains under strict serving constraints; and sequence-level value optimization provides limited credit assignment to intermediate SID decisions, causing likelihood-based beam search to prematurely prune potentially high-value candidates.

We present GR4AD-X, which addresses these limitations through reliable SID construction, serving-efficient capacity scaling, and unified value alignment. Specifically, GR4AD-X improves SID reliability through Multi-Path assignment. It expands model capacity along multiple complementary dimensions rather than relying solely on wider or deeper backbones, achieving a better effectiveness-efficiency trade-off under strict serving constraints. Finally, Unified Value-Guided Generation and Discrimination (UniVGD) aligns generation with commercial value through prefix-level reward optimization, value-guided beam pruning, and discriminative reranking within a unified model.

Large-scale online A/B tests on the Kuaishou advertising platform demonstrate that GR4AD-X achieves a cumulative revenue improvement of \(3.07\%\) over the deployed GR4AD baseline while satisfying strict latency and throughput constraints.

\end{abstract}

%% 生成式推荐+3



\maketitle
% Title-page contribution footnotes; do not change the body footnote counter.
\begingroup
  \renewcommand{\thefootnote}{\fnsymbol{footnote}}
  \footnotetext[1]{These authors contributed equally.}
  \footnotetext[2]{Corresponding author.}
\endgroup

% \vspace{-6pt}
\section{Introduction}
\label{sec:introduction}

% GR4AD~\cite{xue2026gr4ad} established a generative paradigm for large-scale advertising recommendation. It represents advertisements with unified semantic identifiers (SIDs), generates candidates autoregressively, aligns learning with advertising value, and supports real-time deployment through LazyAR and dynamic beam search. These designs replaced fragmented retrieval components with a unified generative model and delivered substantial online gains under strict serving constraints.

GR4AD~\cite{xue2026gr4ad} established a generative production paradigm for Kuaishou's large-scale advertising recommendation, serving over 400 million users. It unifies SID-based advertisement representation, autoregressive candidate generation, value-aware learning, and real-time serving with LazyAR and dynamic beam search. By integrating these functions into a unified generative retrieval framework, GR4AD achieved substantial online revenue gains while satisfying strict latency and throughput constraints.

% Continued production deployment nevertheless reveals three limitations. First, deterministic SID assignment is sensitive to semantic boundaries and inventory drift. Second, conventional backbone expansion yields diminishing returns while increasing autoregressive serving costs. Third, sequence-level value optimization provides limited supervision for intermediate SID decisions, leaving generation probability insufficiently aligned with commercial value. GR4AD-X addresses these limitations through reliable SID construction, serving-efficient capacity scaling, and value alignment throughout generation.

Despite these successes, production deployment reveals further opportunities to improve GR4AD in representation reliability, model scaling, and value-oriented generation. First, deterministic SID assignment remains sensitive to semantic boundaries. Second, conventional model scaling yields diminishing revenue gains under strict online latency constraints. Third, sequence-level value optimization provides limited supervision for intermediate SID decisions, leaving generation probabilities insufficiently aligned with commercial value. GR4AD-X addresses these challenges through reliable SID construction, serving-efficient capacity scaling, and value alignment throughout the generation process.

\begin{figure}[t]
  \centering
  \includegraphics[width=\linewidth]{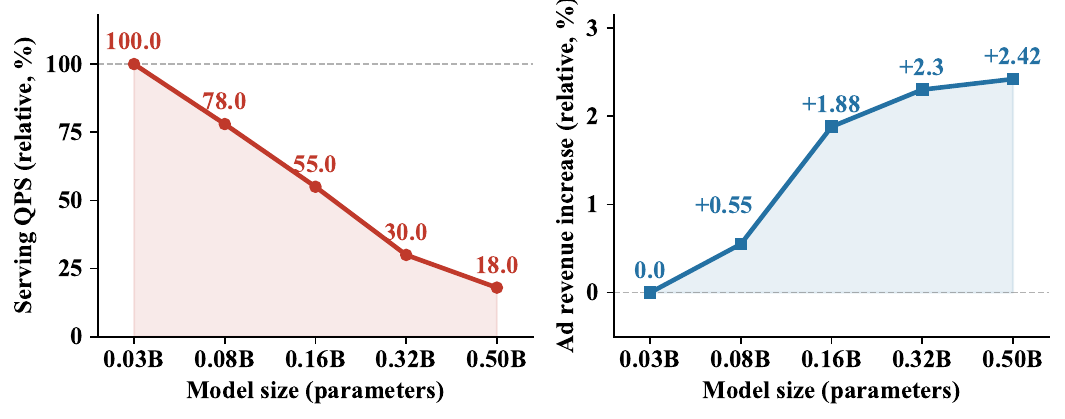}
    \caption{Production trade-off of naive backbone scaling.
    All variants use Multi-Path SID. Serving QPS and revenue
    improvements are reported relative to the $0.03$B variant.
    Larger backbones substantially reduce serving throughput,
    while revenue gains exhibit diminishing returns.}
  \label{fig:naive-scaling}
  \vspace{-14pt}
\end{figure}

\noindent\textbf{Reliable SID representation.}
The first SID token determines the coarse semantic partition and
constrains subsequent decisions. For advertisements near
partition boundaries, hard quantization forces one of several
plausible assignments, introducing noise at this critical step
and making the advertisement unreachable if its assigned prefix
is pruned during beam search. We propose Multi-Path SID to
retain the top-$m$ first-level assignments and complete each
with a separate residual quantization trajectory, providing
alternative generation paths for retrieving the advertisement.

% \textbf{Serving-efficient capacity scaling.}
% Conventional scaling widens hidden dimensions or adds more model layers, increasing the cost of every beam-decoding step. In production, expanding the backbone from $0.16$B to $0.50$B parameters yields only a $0.30\%$ revenue improvement while nearly doubling inference resource usage. Figure~\ref{fig:naive-scaling} shows that the gain quickly saturates as serving QPS declines. GR4AD-X instead scales input signals, encoder capacity, vocabulary memory, residual streams, and generation decision depth. This design increases effective capacity without proportionally increasing dense backbone computation.

\noindent\textbf{Serving-efficient capacity scaling.}
Scaling model capacity is a natural direction for improving GR4AD, as LLMs can better capture heterogeneous advertising features and diverse user behavior patterns. However, naive parameter scaling is inefficient under the strict latency and throughput constraints of production advertising systems. As shown in Figure~\ref{fig:naive-scaling}, increasing model size substantially reduces serving QPS, while the corresponding revenue gains quickly become marginal. GR4AD-X therefore revisits the scaling strategy and expands capacity along five complementary dimensions: input signals, encoder capacity, vocabulary memory, residual streams, and generation decision depth. This design improves effective model capacity with only modest serving overhead, providing a more efficient scaling path than enlarging the backbone.

\noindent\textbf{Value-aligned generation.}
Next-token prediction fits logged interactions, whereas advertising systems optimize commercial value such as eCPM. In our production diagnosis, HitRate@10 between the advertisements with the highest eCPM and those ranked highest by generation probability is only approximately $0.02$. Low overlap at different SID prefix levels indicates that this mismatch persists throughout generation. Under beam search, an underestimated prefix may cause a high-eCPM advertisement to be pruned before its complete SID is generated. Sequence-level optimization provides insufficient credit assignment to identify the responsible prefix decision. We therefore introduce Unified Value-Guided Generation and Discrimination (UniVGD). PV-RSPO propagates eCPM rewards through the SID prefix tree before generation, PV-BeamSearch incorporates prefix values into beam pruning during generation, ensures the generation process does not rely solely on generation probabilities, and discriminative heads rerank complete candidates after generation. These components share user representations and are jointly optimized within one model.

Overall, GR4AD-X makes the following contributions.
\begin{itemize}[leftmargin=*,nosep]
  % \item We develop reliable SID construction through multi-path assignment, power-diagram quantization, and temporally stratified codebook learning, reducing assignment ambiguity while maintaining consistency between offline construction and online insertion.
  \item We propose Multi-Path SID, which retains multiple
    first-level assignments and continues residual quantization
    along each branch. This provides alternative generation paths for advertisements near partition boundaries, improving their opportunities to survive beam pruning without increasing decoding depth.

  \item We design a serving-efficient capacity scaling strategy that expands model capacity along input signals, encoder capacity, vocabulary memory, residual streams, and generation decision depth, achieving a better effectiveness-efficiency trade-off than naive model scaling.

  \item We introduce UniVGD as a unified value alignment framework, integrating prefix-level eCPM optimization, value-guided beam pruning, and discriminative reranking across the complete generation process.
\end{itemize}
\section{Related Work}
\subsection{GR and Semantic Item Identifiers}
Generative recommendation models retrieval as autoregressive prediction over discrete item identifiers. Building on TIGER's residual-quantized semantic IDs (SIDs)~\cite{rajput2023tiger}, subsequent methods improve collaborative semantics, tokenization, hierarchical structure, and generalization~\cite{zheng2024lcrec,wang2024colarec,si2024seater,wang2024letter,qu2025tokenrec}. Recent studies examine balanced SID construction, multimodal quantization, and systematic evaluation~\cite{ju2025handbook,fu2025forge,he2025plum,luo2025qarm,xu2025mmq}, while industrial systems validate generative retrieval in recommendation, advertising, and e-commerce search~\cite{zhou2025onerec,zheng2025egav2,xue2026gr4ad,zhang2025gpr,zhou2025openonerec,chen2025onesearch}. Most methods still assign each item a single path and remain vulnerable to early quantization errors~\cite{baikalov2026staleness}. Multi-Path SID improves first-level assignment reliability with alternative retrieval paths for ads near semantic boundaries.
\subsection{Value Alignment in GR}
Next-token prediction fits logged interactions but does not optimize advertising value. Industrial systems introduce value-aware supervision, preference optimization, hierarchical rewards, or reward-weighted ranking~\cite{zheng2025egav2,xue2026gr4ad,zhang2025gpr,chen2025onesearch}. Recent methods guide SID construction and decoding with value signals~\cite{jiang2026vstar,zhang2026univa}, while BEAR addresses premature prefix pruning caused by training–inference mismatch~\cite{yang2026bear}. UniVGD instead coordinates the full generation process through eCPM reward propagation to prefixes, value-guided beam pruning, and candidate reranking.
\subsection{Model Scaling in GR and LLMs}
Industrial generative recommenders must scale under strict latency and throughput constraints. Recommendation scaling typically enlarges input sequences, hidden dimensions, or Transformer depth~\cite{zhou2025onerec,xue2026gr4ad,he2025plum,zhou2025openonerec}, while test-time methods expand decoding budgets through wider beam search and parallel generation~\cite{hou2025rpg,wang2025nezha}. Although effective, these approaches substantially increase online latency, making naive scaling inefficient for production advertising systems.
Similar constraints have motivated efficient LLM scaling through vocabulary or memory expansion~\cite{cheng2026engram,sadhukhan2026stem,wu2026msn}, chain-of-thought decision depth~\cite{wei2022chain}, and multi-stream residual pathways~\cite{xie2025mhc}. Inspired by these directions, GR4AD-X jointly scales input context, encoder capacity, vocabulary memory, residual streams, and decision depth under a fixed serving budget, improving the effectiveness-efficiency trade-off over naive model scaling.
\section{Method}
\subsection{Overview}
\begin{figure*}[t]
  \centering
  \includegraphics[width=\linewidth]{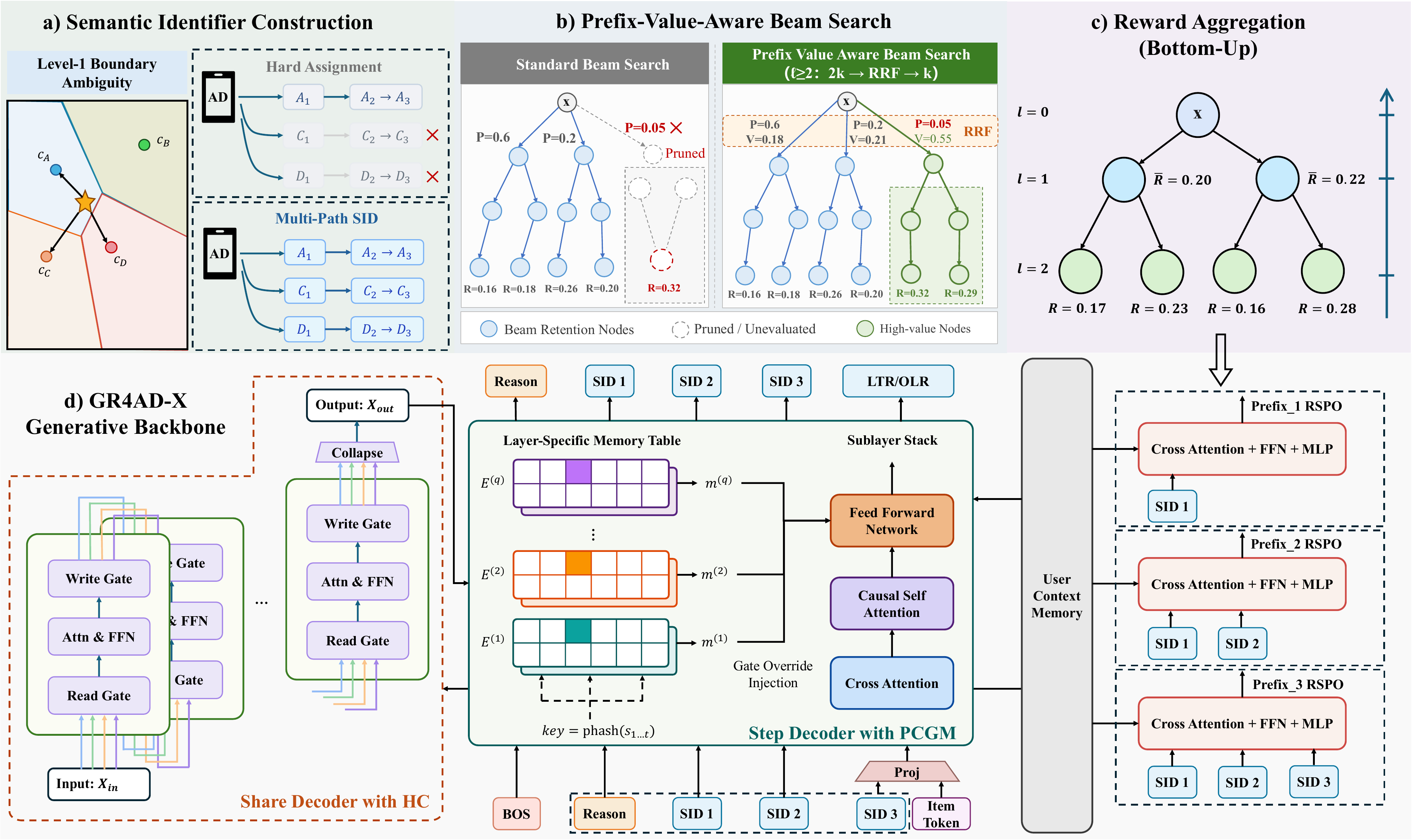}
  \caption{Overview of GR4AD-X. \textbf{(a)} Multi-Path SID retains multiple plausible first-level assignments for boundary ads. \textbf{(b)} PV-BeamSearch expands to $2k$ candidates and fuses generation and prefix-value ranks via RRF. \textbf{(c)} Revenue rewards propagate bottom-up through the SID prefix tree for PV-RSPO. \textbf{(d)} The backbone uses a Shared Decoder with diagonal Hyper-Connections, a Step Decoder with PCGM, and per-level prefix-value heads for value alignment and discriminative reranking.
}
  \label{fig:overview}
  \vspace{-6pt}
\end{figure*}

% Figure~\ref{fig:overview} presents the overall architecture of
% GR4AD-X. Multi-Path SID retains multiple first-level assignments
% and completes each with a separate residual quantization
% trajectory. These paths serve as joint supervision targets,
% allowing an advertisement to be retrieved through alternative
% SID prefixes (Section~\ref{sec:sid}). Serving-efficient capacity
% scaling improves the model along five complementary dimensions:
% input signals, encoder capacity, vocabulary memory, residual
% streams, and generation decision depth
% (Section~\ref{sec:capacity}). Finally, UniVGD aligns generation
% with commercial value through prefix-level reward optimization
% during pre-generation, value-guided beam pruning during
% in-generation, and discriminative reranking during
% post-generation (Section~\ref{sec:univgd}).

Figure~\ref{fig:overview} summarizes the architecture of GR4AD-X. Multi-Path SID retains multiple first-level assignments, each completed with a separate residual quantization trajectory. Joint supervision on these paths enables advertisement retrieval through alternative SID prefixes (Section~\ref{sec:sid}). Serving-efficient capacity scaling targets five complementary dimensions: input signals, encoder capacity, vocabulary memory, residual streams, and generation decision depth (Section~\ref{sec:capacity}). Finally, UniVGD aligns generation with commercial value through pre-generation prefix-level reward optimization, in-generation value-guided beam pruning, and post-generation discriminative reranking (Section~\ref{sec:univgd}).

\subsection{Multi-Path Semantic Identifiers}
\label{sec:sid}

Semantic identifiers (SIDs) define both the supervision targets and
the discrete search space of a generative recommender. A conventional
single-path SID associates each advertisement with one hierarchical
quantization trajectory. For advertisements near coarse cluster
boundaries, however, multiple first-level centroids may be similarly
close in the embedding space. Restricting such an advertisement to a
single first-level prefix makes it unreachable whenever that prefix
is pruned during decoding. We introduce Multi-Path SID to provide
alternative retrieval paths while preserving the residual
quantization structure.

\noindent\textbf{Residual quantization.}
Given the semantic embedding $\mathbf e_i\in\mathbb R^d$ of
advertisement $i$, let
$\mathcal C_l=\{\mathbf c_{l,k}\}_{k=1}^{K_l}$ denote the codebook
at level $l$. Standard RQ-KMeans initializes
$\mathbf r_i^{(1)}=\mathbf e_i$ and recursively computes
\begin{equation}
\begin{aligned}
s_l
&=
\operatorname*{\arg\min}_{k\in\{1,\ldots,K_l\}}
\left\|\mathbf r_i^{(l)}-\mathbf c_{l,k}\right\|_2^2,\\
\mathbf r_i^{(l+1)}
&=
\mathbf r_i^{(l)}-\mathbf c_{l,s_l},
\qquad l=1,\ldots,L.
\end{aligned}
\label{eq:rq}
\end{equation}
This produces a single SID sequence $S_i=(s_1,\ldots,s_L)$.
Multi-Path SID modifies only the first-level selection and uses
the same codebooks for all subsequent residual trajectories.

\noindent\textbf{Multi-path construction.}
Instead of selecting one first-level centroid, we retain the
$m$ nearest centroids:
\begin{equation}
\left(s_1^{(1)},\ldots,s_1^{(m)}\right)
=
\operatorname*{top\text{-}m}_{k}
\left(-\left\|\mathbf e_i-\mathbf c_{1,k}\right\|_2^2\right).
\label{eq:multipath-topm}
\end{equation}
Each selected centroid initializes a separate residual branch:
\begin{equation}
\mathbf r_i^{(2,j)}
=
\mathbf e_i-\mathbf c_{1,s_1^{(j)}},
\qquad j=1,\ldots,m.
\label{eq:multipath-residual}
\end{equation}
For each branch, we apply Eq.~\eqref{eq:rq} from level $2$
to level $L-1$. Following GR4AD, the final level uses a
hash-based mapping of non-semantic advertisement features
rather than residual quantization. Denoting this final token
by $h_i$, the complete SID paths are
\begin{equation}
\begin{aligned}
S_i^{(j)}
&=\left(s_1^{(j)},\ldots,s_{L-1}^{(j)},h_i\right),\\
\mathcal S_i
&=\left\{S_i^{(j)}\right\}_{j=1}^{m}.
\end{aligned}
\label{eq:multipath-sequences}
\end{equation}
The hash mapping is unchanged from GR4AD, and all paths
of the same advertisement share the final token
$s_L^{(j)}=h_i$.
Branching occurs only at the first semantic level:
each subsequent quantization level selects one centroid
per branch. Consequently, each advertisement receives
$m$ length-$L$ paths rather than an exponentially expanding
set of paths. Setting $m=1$ recovers the single-path SID
construction used in GR4AD.

\noindent\textbf{Blockwise multi-target training.}
For a positive advertisement $i$ and user context $x$, we train
the model on all paths in $\mathcal S_i$. The objective is
\begin{equation}
\mathcal L_{\mathrm{SID}}
=
-\sum_{j=1}^{m}\alpha_j
\sum_{l=1}^{L}
\log\pi_\theta
\left(s_l^{(j)}\mid x,S_{<l}^{(j)}\right),
\label{eq:multi-sid-loss}
\end{equation}
where $S_{<l}^{(j)}=(s_1^{(j)},\ldots,s_{l-1}^{(j)})$,
$\alpha_j\geq 0$, and $\sum_{j=1}^{m}\alpha_j=1$.

To process multiple targets in one forward pass, we pack the
teacher-forced inputs for all paths into separate SID blocks.
A blockwise causal attention mask prevents each block from
attending to tokens in other blocks, while preserving access
to the shared user context and its own preceding SID tokens.
This avoids duplicating the user-context computation across
paths while retaining path-specific autoregressive supervision.

\noindent\textbf{Inference and deduplication.}
At inference time, the model generates SID sequences using the
standard autoregressive decoding procedure. Multi-path indexing
allows different generated sequences to resolve to the same
advertisement, providing alternative first-level prefixes through
which it can be retrieved.

Because multiple paths may retrieve the same advertisement,
we deduplicate candidates at the advertisement level and retain
the highest-scoring path for each advertisement. Although multiple paths may consume additional beam slots, advertisement-level deduplication prevents repeated paths from appearing as duplicate items in the final candidate set. The net effectiveness of this design is evaluated through production A/B tests.

\subsection{Efficient Scaling for Real-Time Serving}
\label{sec:capacity}

GR4AD must scale under strict latency and hardware constraints, making naive model scaling inefficient. GR4AD-X therefore adopts serving-efficient capacity scaling along five dimensions: input signals, encoder capacity, vocabulary memory, residual streams, and generation decision depth. 

\subsubsection{Input and Encoder Scaling}
We enrich input signals with over 20 user-profile and contextual features from
the production ranking system, reallocating sequence budget by feature
importance to keep the input length unchanged. We also enhance encoder
nonlinearity by increasing the embedding dimensions of key high-cardinality features from 16 to 64
and replacing the conventional FFN with a SwiGLU-based FFN, improving user
representation with marginal serving overhead.

\subsubsection{Vocabulary Scaling with Prefix-Conditioned Gated Memory}
\label{sec:pcgm}

SID embeddings act as token-indexed parametric memory that stores
advertisement semantics and user-advertisement collaborative signals.
Expanding this memory provides an efficient way to increase model capacity
without proportionally increasing the dense computation activated at each
decoding step. Prefix-Conditioned Gated Memory (PCGM) scales vocabulary memory along two complementary axes.

\noindent\textbf{Vertical scaling: layer-specific memory tables.}
We allocate an independent SID memory table \(E^{(q)}\) for each decoder layer
\(q\), allowing every layer to directly access the original SID-indexed
information. The retrieved memory is injected through the gated FFN, so each
layer can adaptively determine how much raw SID information to incorporate.
Here, \(q\) denotes the decoder-layer index and is distinct from the SID-level
index \(l\).

\noindent\textbf{Horizontal scaling: SID prefix n-gram memory.}
SIDs are hierarchical codes whose prefixes capture semantics at
different granularities. We construct memory entries for 1-gram,
2-gram, and 3-gram SID prefixes. Partial-prefix memories support
the prediction of subsequent SID tokens, while complete-SID
memory supports candidate scoring in the rank step. Since the
complete SID is nearly item-specific, its memory can store
fine-grained advertisement semantics and collaborative signals.

We illustrate the polynomial hashing scheme with the 3-gram case:
\begin{equation}
  \mathrm{phash}(s_1,s_2,s_3)
  =
  \left(s_1\rho^2+s_2\rho+s_3\right)\bmod M,
  \label{eq:pcgm-hash}
\end{equation}
where $\rho$ is a fixed prime and $M$ controls the table size.
Lower-order prefixes are handled analogously.

\noindent\textbf{Prefix-conditioned gating.}
PCGM retrieves memory using only SID tokens already available
at the current decoding position. The first SID token $s_1$ is
predicted without PCGM. After $s_1$ is generated, its 1-gram
memory is used when predicting $s_2$; the 2-gram memory of
$(s_1,s_2)$ is then used when predicting $s_3$. Once the complete
SID is available, its 3-gram memory is used in the rank step.
For a prefix of length $t$, the lookup key is
\begin{equation}
  k_t =
  \begin{cases}
    s_1, & t=1,\\
    \mathrm{phash}(s_1,s_2), & t=2,\\
    \mathrm{phash}(s_1,s_2,s_3), & t=3.
  \end{cases}
  \label{eq:pcgm-key}
\end{equation}
Here, $t$ denotes the length of the available SID prefix,
not the index of the token being predicted. Each decoder
layer $q$ retrieves its own memory vector:
\begin{equation}
  \mathbf m_t^{(q)}=E^{(q)}[k_t].
  \label{eq:pcgm-lookup}
\end{equation}
The retrieved vector replaces the dense FFN gate:
\begin{equation}
  \mathrm{FFN}^{(q)}
  \left(\mathbf h_t^{(q)},\mathbf m_t^{(q)}\right)
  =
  W_d^{(q)}
  \left(
    W_u^{(q)}\mathbf h_t^{(q)}
    \odot
    \mathbf m_t^{(q)}
  \right),
  \label{eq:pcgm-ffn}
\end{equation}
where $\mathbf h_t^{(q)}$ denotes the hidden state at the
position using the length-$t$ prefix memory.

In GR4AD, the complete-SID position was used to predict an
eCPM token. GR4AD-X instead incorporates eCPM information into
the advertisement-side features of the rank token and performs
discriminative candidate scoring
(Section~\ref{sec:univgd:post}). Accordingly, the complete-SID
memory is used in the rank step rather than for a separate
eCPM-token prediction.

\subsubsection{Multi-Stream Residuals with Diagonal Hyper-Connections}
\label{sec:diagonal-hc}

A single residual stream must carry heterogeneous user and contextual
signals, which may be overwritten or diluted as depth increases.
We introduce diagonal Hyper-Connections in the Shared Decoder to
expand residual information pathways without replicating the
attention and FFN computation.

Given the input state
$X_{\mathrm{in}}\in\mathbb R^{B\times T\times D}$, we duplicate
it into $n=4$ parallel streams:
\begin{equation}
  Y=\mathrm{Expand}(X_{\mathrm{in}})
  \in\mathbb R^{B\times T\times n\times D},
\end{equation}
where $B$, $T$, and $D$ denote batch size, Shared Decoder sequence
length, and hidden dimension, respectively.

Inspired by the diagonal parameterization in Hyperloop
Transformers~\cite{zeitoun2026hyperloop}, we use input-dependent
gates to read from, write to, and preserve the parallel streams.
For clarity, we omit batch and sequence indices in the gate
definitions below. The read and write operations are applied
position-wise, while attention operates on the full sequence
of read representations. Each sublayer follows a
read–compute–write cycle:
\begin{equation}
\begin{aligned}
  \mathbf x_{\mathrm{read}} &= H_{\mathrm{pre}}Y,\\
  Y' &= H_{\mathrm{res}}Y
       +H_{\mathrm{post}}\,
        \mathrm{Sublayer}(\mathbf x_{\mathrm{read}}),
\end{aligned}
\label{eq:hc-update}
\end{equation}
where $\mathrm{Sublayer}$ denotes an attention or FFN sublayer,
$H_{\mathrm{pre}}\in\mathbb R^{1\times n}$,
$H_{\mathrm{post}}\in\mathbb R^{n\times1}$, and
$H_{\mathrm{res}}\in\mathbb R^{n\times n}$ is diagonal.
Thus, each stream receives a scalar gate shared across its
$D$ channels, while each sublayer operates on a single
$D$-dimensional representation.

The gates are computed from the normalized multi-stream state
$\mathbf z=\mathrm{RMSNorm}(\mathrm{flatten}(Y))
\in\mathbb R^{nD}$:
\begin{equation}
\begin{aligned}
  \mathbf g_{\mathrm{pre}}
    &= \sigma\!\left(
       \alpha_{\mathrm{pre}}W_{\mathrm{pre}}\mathbf z
       +\mathbf b_{\mathrm{pre}}\right),\\
  \mathbf g_{\mathrm{post}}
    &= 2\sigma\!\left(
       \alpha_{\mathrm{post}}W_{\mathrm{post}}\mathbf z
       +\mathbf b_{\mathrm{post}}\right),\\
  \mathbf g_{\mathrm{res}}
    &= \sigma\!\left(
       \alpha_{\mathrm{res}}W_{\mathrm{res}}\mathbf z
       +\mathbf b_{\mathrm{res}}\right),
\end{aligned}
\label{eq:hc-gates}
\end{equation}
with
\begin{equation}
  H_{\mathrm{pre}}=\mathbf g_{\mathrm{pre}}^\top,
  \qquad
  H_{\mathrm{post}}=\mathbf g_{\mathrm{post}},
  \qquad
  H_{\mathrm{res}}=\mathrm{diag}(\mathbf g_{\mathrm{res}}).
\end{equation}
For each gate type $a\in\{\mathrm{pre},\mathrm{post},\mathrm{res}\}$,
$W_a\in\mathbb R^{n\times nD}$,
$\mathbf b_a\in\mathbb R^n$, and $\alpha_a\in\mathbb R$
are learnable parameters.

A learned aggregation collapses the final multi-stream state
back to $X_{\mathrm{out}}\in\mathbb R^{B\times T\times D}$.

\subsubsection{Semantically Anchored Reason Prefill}
\label{sec:sarp}

The first-level SID $s_1$ determines the coarse branch of the
generation path, yet must be predicted without any preceding
SID token. We insert a continuous reasoning token between BOS
and $s_1$, providing an intermediate computation position
before this decision.

Given user context $x$, a learnable network $f_r$ maps the BOS
hidden state to the reasoning-token embedding:
\begin{equation}
  \mathbf t^{\mathrm{reason}}(x)
  =
  f_r\left(\mathbf h^{\mathrm{BOS}}(x)\right).
  \label{eq:reason-token}
\end{equation}
The decoder processes this embedding to obtain
$\mathbf h^{\mathrm{reason}}(x)$, which is shared across
all retained SID paths. This state is used to predict
their first-level tokens and receives auxiliary coarse
semantic supervision. It is computed from the user context
and BOS before any target SID token is observed, and is
therefore independent of the chosen SID path.

To anchor this state in SID semantics, we recluster the frozen
first-level RQ-KMeans codebook
$\mathcal C=\{\mathbf c_k\}_{k=1}^{K}$ into a smaller codebook
$\mathcal C'=\{\mathbf c'_v\}_{v=1}^{K'}$, where $K'\ll K$.
The nearby first-level centroids retained by Multi-Path SID
map to the same coarse prototype in most cases.
We therefore use a single advertisement-level coarse label
$a_i$ to supervise the reasoning state shared across paths,
while the multi-target SID loss preserves their distinct
fine-grained generation targets.

We project the decoder hidden state into the coarse codebook
space and normalize both the projection and the prototypes:
\begin{equation}
  \mathbf q(x)
  =
  \frac{W_r\mathbf h^{\mathrm{reason}}(x)}
       {\|W_r\mathbf h^{\mathrm{reason}}(x)\|_2},
  \qquad
  \widehat{\mathbf c}'_v
  =
  \frac{\mathbf c'_v}{\|\mathbf c'_v\|_2}.
  \label{eq:reason-projection}
\end{equation}
Prototype classification supervises the target coarse cluster:
\begin{equation}
  \mathcal L_{\mathrm{proto}}(x,i)
  =
  -\log
  \frac{
    \exp\left(
      \mathbf q(x)^\top\widehat{\mathbf c}'_{a_i}/\tau
    \right)
  }{
    \displaystyle\sum_{v=1}^{K'}
    \exp\left(
      \mathbf q(x)^\top\widehat{\mathbf c}'_v/\tau
    \right)
  },
  \label{eq:reason-prototype}
\end{equation}
where $\tau$ is the temperature. A cosine loss further
encourages alignment with the target prototype:
\begin{equation}
  \mathcal L_{\mathrm{cos}}(x,i)
  =
  1-\mathbf q(x)^\top\widehat{\mathbf c}'_{a_i}.
  \label{eq:reason-cosine}
\end{equation}
The combined auxiliary loss is
\begin{equation}
  \mathcal L_{\mathrm{reason}}(x,i)
  =
  \mathcal L_{\mathrm{proto}}(x,i)
  +
  \lambda_{\mathrm{cos}}\mathcal L_{\mathrm{cos}}(x,i),
  \label{eq:reason-loss}
\end{equation}
where $\lambda_{\mathrm{cos}}$ balances the two objectives.

At inference time, the reasoning token is generated from the
BOS state without requiring a target advertisement or coarse
label. SARP thus adds one semantically supervised decoding
position before first-level SID prediction while preserving
the SID depth.

\subsection{UniVGD: Unified Value-Guided Generation and Discrimination}
\label{sec:univgd}

In advertising recommendation systems, the ranking order and commercial value of generated candidates are directly tied to platform revenue and the fulfillment of advertiser objectives. To this end, we pursue a more thorough optimization of value perception throughout the system. We propose UniVGD, a framework that incorporates discriminative value awareness at each stage of the generation pipeline to guide the ultimate decision-making process. 
% It is built upon a shared user memory:
% \emph{(i)} \textbf{pre-generation}, PV-RSPO propagates an item-level reward into prefix-level targets and performs level-wise ranking-aligned optimization; \emph{(ii)} \textbf{in-generation}, PV-BeamSearch lets
% discriminative evidence participate in the pruning rule itself; \emph{(iii)} \textbf{post-generation}, a rank step
% re-orders the decoded set with calibrated discriminative heads. The first two govern \emph{reachability}, the third governs \emph{order}.

\subsubsection{Pre-Generation: Prefix-Value-Aware RSPO}
\label{sec:univgd:pre}

Sequence-level rewards assign the same credit to all SID decisions,
penalizing a useful coarse prefix even when only a later token is wrong. This creates a
credit-assignment mismatch for hierarchical SID decoding: the action space is
a tree, while the reward lives only at its leaves. We address this mismatch
by aggregating leaf rewards bottom-up into a value for every prefix and then
learning these values with one lightweight discriminative head per level.

Using the Multi-Path notation in Section~\ref{sec:sid}, let
\(p=S_{i,1:l}^{(j)}\) be advertisement \(i\)'s level-\(l\) prefix on path
\(j\). For user \(u\), let \(\mathcal R(u)\) be the rolled-out advertisements,
\(r(u,i)\) their rewards, and
\(\mathcal S_u^{(j)}(p)=\{i\in\mathcal R(u):S_{i,1:l}^{(j)}=p\}\).

% \begin{equation}
% \mathcal S_u(p)
% =
% \left\{
% i\in\mathcal R(u):
% S_{i,1:l}=p
% \right\}
% \label{eq:prefix-descendants}
% \end{equation}

\noindent\textbf{Bottom-up reward aggregation.}
A prefix receives the mean reward of its sampled descendants:
\begin{equation}
V^{(j)}(u,p)
=
\frac{1}{|\mathcal S_u^{(j)}(p)|}
\sum_{i\in\mathcal S_u^{(j)}(p)}
r(u,i),
\qquad
|\mathcal S_u^{(j)}(p)|>0.
\label{eq:prefix-value}
\end{equation}
Equivalently, values propagate bottom-up in the prefix tree
\(\mathcal T^{(j)}\) formed by the \(j\)-th paths:
\begin{equation}
V^{(j)}(u,p)
=
\frac{
\sum_{v\in\mathrm{ch}^{(j)}(p)}
|\mathcal S_u^{(j)}(v)|V^{(j)}(u,v)
}{
\sum_{v\in\mathrm{ch}^{(j)}(p)}
|\mathcal S_u^{(j)}(v)|
},
\qquad
V^{(j)}(u,S_i^{(j)})=r(u,i).
\label{eq:recursive}
\end{equation}
Here, \(\mathrm{ch}^{(j)}(p)\) denotes the children of \(p\).

\noindent\textbf{Lightweight prefix-level heads.}
Each SID level has a lightweight head \(g_l(\cdot;\theta_l)\), shared
across paths and attending to the backbone's user memory \(\mathbf M_u\):
\begin{equation}
\widehat V_l(u,p)
=
g_l
\left(
\mathrm{CrossAttn}
\left(
\mathbf e_p,\mathbf M_u
\right)
\right),
\qquad
\mathbf e_p=\mathrm{Emb}(p).
\label{eq:head}
\end{equation}
The listwise RSPO loss~\cite{xue2026gr4ad} is applied separately to each
level and path, then aggregated with the weights from
Eq.~\eqref{eq:multi-sid-loss}:
\begin{equation}
\begin{aligned}
\mathcal L_{\mathrm{PV\text{-}RSPO}}
&=
\sum_{j=1}^{m}\alpha_j
\mathcal L_{\mathrm{PV\text{-}RSPO}}^{(j)},\\
\mathcal L_{\mathrm{PV\text{-}RSPO}}^{(j)}
&=
\sum_{l=1}^{L}
\beta_l
\mathcal L_{\mathrm{RSPO}}
\left(
\{\widehat V_l(u,p)\}_{p\in\mathcal P_l^{(j)}(u)},
\{V^{(j)}(u,p)\}_{p\in\mathcal P_l^{(j)}(u)}
\right).
\end{aligned}
\label{eq:pv-rspo}
\end{equation}
Here, \(\mathcal P_l^{(j)}(u)\) contains the distinct level-\(l\) prefixes
of the \(j\)-th paths in \(\mathcal R(u)\), and \(\beta_l\) weights each level.

\subsubsection{In-Generation: Prefix-Value-Aware Beam Search}
\label{sec:univgd:in}

Section~\ref{sec:univgd:pre} injects prefix-level value into training. At
inference time, however, standard beam search still expands and prunes purely
by generation probability. A branch with low shallow-level likelihood
but high downstream reward may therefore be discarded
irreversibly. Once a high-value prefix falls out of the beam at level \(l\),
no later evidence can bring it back. PV-BeamSearch reuses the prefix-value
heads as a second voter during expansion, turning pruning from a
likelihood-only decision into a likelihood-value consensus.

Let \(\mathcal B_{l-1}\) denote the beam after level \(l-1\), and let \(k\)
denote the beam width. For every level \(l\ge2\), we first retain a relaxed
candidate pool \(\mathcal A_l\) containing the top-\(2k\) expansions ranked
by generation probability \(\log\pi_\theta(p)\). We then score the same
candidates with the predicted prefix value \(\widehat V_l(u,p)\) and fuse the
two rankings through Reciprocal Rank Fusion:
\begin{equation}
\mathrm{RRF}(p)
=
\frac{1}{\kappa+\mathrm{rank}_{\pi}(p)}
+
\frac{1}{\kappa+\mathrm{rank}_{V}(p)},
\label{eq:rrf-score}
\end{equation}
\begin{equation}
\mathcal B_l
=
\operatorname*{top\text{-}k}_{p\in\mathcal A_l}
\mathrm{RRF}(p),
\label{eq:rrf}
\end{equation}
where \(\kappa\) is a smoothing constant. Rank-based fusion is preferred over
score interpolation because \(\log\pi_\theta(p)\) and
\(\widehat V_l(u,p)\) are not calibrated on a common scale. The widened
candidate pool lets the value head nominate prefixes that
generation probability alone would have removed, while rank fusion prevents
it from overriding generation quality outright. Figure~\ref{fig:overview}(b)
illustrates this process.

\noindent\textbf{Why fusion starts at \(l\ge2\).}
Beam search is a prefix-dependent process, and the first-level SID determines
the coarse region of the SID space in which all subsequent expansions are
legal. In our experiments, applying discriminative fusion at level \(1\)
substantially degraded the validity of decoded candidates because the value
head could nominate coarse codes under which few well-formed advertisements
exist. We therefore leave the first level untouched and intervene only after
the hypothesis is anchored.

\subsubsection{Post-Generation: Discriminative Re-Ranking}
\label{sec:univgd:post}

The two previous components inject value awareness before and during
generation, yet the final ordering of the beam is still produced by a decoder
trained for next-token prediction. The last stage therefore hands the
ordering decision to an explicitly discriminative scorer. Once beam search
returns the candidate set, a rank step re-scores every candidate and reorders
the set. This scorer is not an external model but an additional step of the
same network, making UniVGD a unified generative-discriminative framework
rather than a generator followed by an independent ranker.

\noindent\textbf{Rank token construction.}
The rank step must consume advertisement-side signals that are unavailable to
the generative branch while remaining compatible with the token interface of
the decoder. For candidate advertisement \(i\) and each path \(j\), let
\(\mathbf z_i^{(j)}=\mathrm{Emb}(S_i^{(j)})\) denote the embedding of its complete SID,
and let
\begin{equation}
\mathbf f_i
=
\left[
\mathbf f_i^{(1)};
\ldots;
\mathbf f_i^{(F)}
\right]
\end{equation}
denote the concatenation of its \(F\) advertisement-side feature fields. These advertisement-side features include the eCPM-related
signals used by the rank step.
The rank token for path \(j\) is
\begin{equation}
\mathbf t_i^{(j)}
=
W_{\mathrm{rk}}
\left[
\mathbf z_i^{(j)};
\mathbf f_i
\right]
+
\mathbf b_{\mathrm{rk}}
\in\mathbb R^d.
\label{eq:rank-token}
\end{equation}
Thus, SID semantics and advertisement-side features are projected into one
vector in the decoder token space. Appending \(\mathbf t_i^{(j)}\) to the
user-side context and running one additional decoding step yields the
path-specific candidate representation
\begin{equation}
\mathbf h_i^{(j)}
=
\mathrm{Decoder}
\left(
\mathbf M_u,
\mathbf t_i^{(j)}
\right),
\label{eq:rank-hidden}
\end{equation}
where \(\mathbf M_u\) is the user-side memory reused without recomputation
from generation. The rank-token projection and decoder parameters are shared across paths, while \(\mathbf f_i\) is the same for all paths of advertisement \(i\).

\noindent\textbf{Discriminative heads.}
Two independent heads are attached to \(\mathbf h_i^{(j)}\), each targeting a
different facet of value, with each head sharing its parameters across SID paths. The value head is trained with the Lambda Loss
objective \(\mathcal L_{\lambda}^{(j)}\) on eCPM, so that the learned ordering is
directly aligned with a position-aware ranking objective. The behavior head
models the progressively deeper conversion chain and is trained with the
neural ordinal logistic regression objective
\(\mathcal L_{\mathrm{OLR}}^{(j)}\), preserving the monotone ordering of
behavior-depth levels rather than treating them as independent binary tasks.
Despite parameter sharing, both losses are computed separately for each SID path and aggregated using the same path weights as in Eq.~\eqref{eq:multi-sid-loss}:
\[
\mathcal L_{\lambda}=\sum_{j=1}^{m}\alpha_j\mathcal L_{\lambda}^{(j)},
\qquad
\mathcal L_{\mathrm{OLR}}=\sum_{j=1}^{m}\alpha_j\mathcal L_{\mathrm{OLR}}^{(j)}.
\]
At inference time, the highest-scoring path retained for each advertisement by the deduplication in Section~\ref{sec:sid} is used for reranking.

Because the two heads optimize different objectives on incomparable scales,
we combine their rankings through Reciprocal Rank Fusion rather than score
interpolation. The final reranking score is
\begin{equation}
\mathrm{Score}(u,i)
=
\mathrm{RRF}
\left(
\mathrm{rank}_{\lambda}(i),
\mathrm{rank}_{\mathrm{OLR}}(i)
\right),
\qquad
i\in\mathrm{Beam}(u).
\label{eq:rerank}
\end{equation}
The complete beam output is reordered according to
\(\mathrm{Score}(u,i)\).

\section{Experiments}

\subsection{Online A/B Tests}

We evaluate GR4AD-X through sequential online A/B tests on the production
advertising system. Starting from the deployed GR4AD baseline, components are
added cumulatively in the order shown in Table~\ref{tab:main-ablation}. Each
row reports the cumulative revenue improvement of the corresponding system
variant relative to the same baseline. 
\begin{table}[t]
\centering
\setlength{\tabcolsep}{8pt}
\renewcommand{\arraystretch}{1.08}
\caption{Sequential online A/B results for GR4AD-X. Components are added
cumulatively, and each row reports the cumulative revenue improvement over
the deployed GR4AD baseline.}
\label{tab:main-ablation}
\begin{tabular}{lc}
\toprule
\textbf{Model Settings} & \textbf{$\Delta$Revenue vs.\ GR4AD} \\
\midrule
\textit{Reliable SID Construction} & \\
\quad + Multi-Path SID                    & $+0.58\%$ \\
\midrule
\textit{Efficient Model Scaling} & \\
\quad + Input \& Encoder Scaling          & $+0.96\%$ \\
\quad + Vocabulary Scaling (PCGM)         & $+1.43\%$ \\
\quad + Stream Scaling (Diag.\ HC)        & $+1.72\%$ \\
\quad + Decision Depth Scaling (SARP)             & $+2.00\%$ \\
\midrule
\textit{Value-Aligned Generation (UniVGD)} & \\
\quad + PV-RSPO                           & $+2.28\%$ \\
\quad + PV-BeamSearch                     & $+2.49\%$ \\
\quad + Discriminative Reranking          & $+3.07\%$ \\
\midrule
\quad \textbf{GR4AD-X (Full)}            & $\mathbf{+3.07\%}$ \\
\bottomrule
\end{tabular}
\vspace{-6pt}
\end{table}

\noindent\textbf{Reliable SID Construction.}
Multi-Path SID improves revenue by $0.58\%$ over the deployed
GR4AD baseline with the same beam width and decoding depth, without increasing inference latency.
This gain highlights the benefit of alternative
retrieval paths under a fixed decoding budget.
Because the first token determines the subsequent generation
subtree, pruning an advertisement's sole first-level prefix
makes it unreachable in later decoding steps.
Retaining multiple plausible first-level paths mitigates
this limitation and improves revenue without changing the backbone.

\noindent\textbf{Efficient Model Scaling.}
Building on Multi-Path SID, the five scaling dimensions increase
the cumulative revenue lift over GR4AD from $0.58\%$ to $2.00\%$,
a gain of $1.42$ percentage points. The consistent gains across
successive additions provide online evidence for the effectiveness
of the proposed scaling strategy. Section~\ref{sec:exp-capacity}
further examines its effectiveness-efficiency trade-off.

\noindent\textbf{Value-Aligned Generation.}
The three UniVGD components increase the cumulative revenue gain from $2.00\%$ to $3.07\%$, contributing an additional $1.07\%$ over the capacity-scaled system. PV-RSPO contributes $0.28\%$ by introducing prefix-level eCPM supervision during training. PV-BeamSearch adds $0.21\%$ by preserving high-eCPM prefixes during decoding, showing that training-time alignment does not eliminate the need for value-aware search. Discriminative reranking provides a further $0.58\%$, the largest individual increment among all evaluated components. This result confirms that fine-grained item-level comparison complements both prefix-level value learning and value-guided generation. Overall, the complete GR4AD-X system achieves a cumulative revenue improvement of $3.07\%$ over the deployed GR4AD baseline.

\subsection{Evaluation of Multi-Path SIDs}
\label{sec:eval-sid}

\begin{figure}[t]
  \centering
  \includegraphics[width=\linewidth]{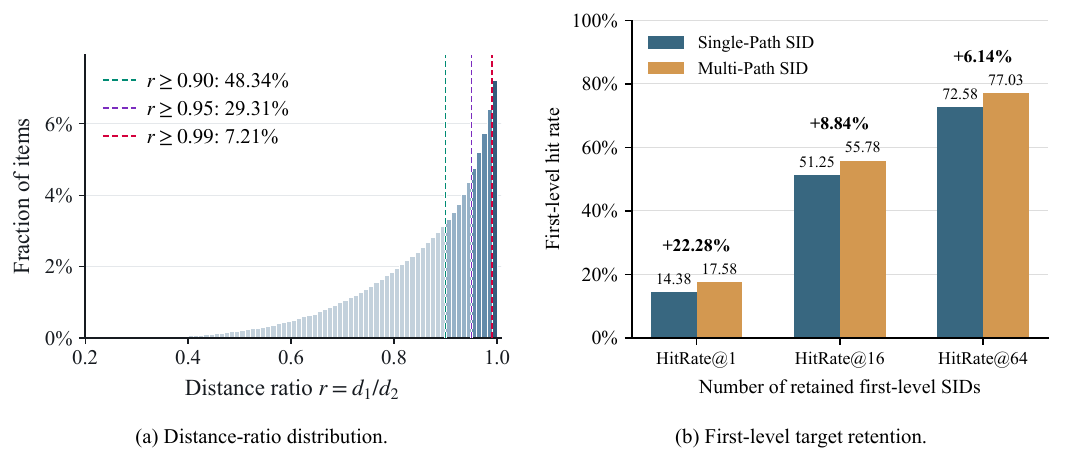}
  \caption{Geometric motivation and first-level target retention
    of Multi-Path SID.
    (a) Distribution of the distance ratio $r=d_1/d_2$
    over 7.556M advertisement embeddings with 16,384
    first-level centroids.
    (b) First-level hit rate of Single-Path and Multi-Path SID
    at $K\in\{1,16,64\}$.
    Labels above each pair indicate relative improvements
    over Single-Path SID.}
  \label{fig:sid-eval}
  \vspace{-12pt}
\end{figure}

We assess Multi-Path SID from two complementary angles:
the geometric structure of first-level centroid assignments
(Figure~\ref{fig:sid-eval}a)
and the resulting improvement in first-level target retention
(Figure~\ref{fig:sid-eval}b).

To quantify the ambiguity of single-path first-level assignments,
we compute the distance ratio
$r = d_1 / d_2$
for each of the 7.556M item embeddings in our corpus,
where $d_1$ and $d_2$ denote the Euclidean distances to the nearest
and second-nearest first-level centroids, respectively.
A ratio close to $1$ indicates that two centroids are nearly
equidistant, making the first-level assignment effectively arbitrary.
Under a codebook of $K_1 = 16{,}384$ first-level centroids,
$48.34\%$, $29.31\%$, and $7.21\%$ of items have $r \geq 0.90$,
$r \geq 0.95$, and $r \geq 0.99$, respectively.
In particular, approximately 2.21M items have $r \geq 0.95$,
meaning their second-nearest centroid is at most $5.26\%$ farther
than their nearest centroid.
For these items, assigning a single first-level prefix is
geometrically fragile: the alternative centroid is nearly as
representative of the item's embedding, yet it is entirely unreachable
under standard RQ-based SIDs.
This mass of ambiguous assignments provides the geometric motivation
for retaining multiple first-level candidates.

We measure first-level Hit@$K$ as the fraction of evaluation
examples for which at least one valid first-level SID token
of the target advertisement appears among the decoder's
top-$K$ tokens, ranked by generation probability given
the user context.
As shown in Figure~\ref{fig:sid-eval}b,
Multi-Path SID improves first-level Hit@$K$
at all evaluated values of $K$.
At $K=1$, the hit rate increases from $14.38\%$ to
$17.58\%$, a relative gain of $22.28\%$.
At $K=16$ and $K=64$, it increases from $51.25\%$
to $55.78\%$ and from $72.58\%$ to $77.03\%$,
respectively, yielding relative gains of $8.84\%$ and $6.14\%$.

These results establish that Multi-Path SID addresses a systematic
coverage gap in single-path quantization: a substantial fraction
of items sit near first-level centroid boundaries, and providing an
alternative path for those items directly increases the probability
that the ground-truth SID prefix survives decoding.

\subsection{Evaluation of Efficient Model Scaling}
\label{sec:exp-capacity}

We evaluate five complementary scaling dimensions through sequential
online A/B tests on the production advertising system. Starting from
the $0.16$B GR4AD model enhanced with Multi-Path SID, we add each
dimension cumulatively in the order shown in
Table~\ref{tab:capacity-ablation}. The table reports revenue
improvements relative to the Multi-Path SID baseline, computed as
$(R/R_{\mathrm{MP}}-1)\times100\%$, where $R$ and $R_{\mathrm{MP}}$
are variant and baseline revenues. In contrast,
Table~\ref{tab:main-ablation} uses the deployed GR4AD system as its
reference. Total parameter counts and normalized serving QPS
are reported alongside revenue to characterize the
effectiveness-efficiency trade-off.

\begin{table}[t]
\centering
\setlength{\tabcolsep}{4pt}
\renewcommand{\arraystretch}{1.05}
\caption{Sequential online A/B results of serving-efficient
capacity scaling. Each proposed dimension is added on top of
the preceding variant. Revenue improvement and QPS are normalized
to the $0.16$B Multi-Path SID baseline. The naive scaling result
is derived from the same Multi-Path SID-enabled experiments in
Figure~\ref{fig:naive-scaling}, re-normalized to the $0.16$B baseline.}
\label{tab:capacity-ablation}
\begin{tabular}{lccc}
\toprule
Variant & $\Delta$Revenue & Parameters & QPS \\
\midrule
Multi-Path SID baseline   & --        & $0.16$B & $100\%$ \\
\midrule
+ Input scaling          & $+0.20\%$ & $0.16$B & $100\%$ \\
+ Encoder scaling        & $+0.38\%$ & $0.20$B & $90\%$  \\
+ Vocabulary scaling     & $+0.85\%$ & $0.46$B & $95\%$  \\
+ Stream scaling         & $+1.13\%$ & $0.49$B & $90\%$  \\
+ Decision-depth scaling & $+1.41\%$ & $0.49$B & $85\%$  \\
\midrule
Naive parameter scaling  & $+0.53\%$ & $0.50$B & $33\%$  \\
\bottomrule
\end{tabular}
\vspace{-16pt}
\end{table}

\noindent\textbf{Overall effectiveness-efficiency trade-off.}
The proposed strategy increases the parameter count from $0.16$B
to $0.49$B, approximately a $3.1\times$ expansion, and delivers
a $+1.41\%$ revenue improvement while retaining $85\%$ of baseline
serving QPS. In comparison, enlarging the dense backbone to
$0.50$B yields only $+0.53\%$ improvement while retaining
$33\%$ QPS. At comparable total parameter counts, the proposed
strategy therefore provides more than twice the relative revenue
improvement and serving throughput of naive backbone scaling.

\noindent\textbf{Per-dimension analysis.}
Input scaling enriches user-profile and contextual signals while
preserving the input sequence length, yielding a $+0.20\%$ revenue
improvement with no change in parameter count or QPS at the
reported precision. Encoder scaling further raises the improvement
to $+0.38\%$, with the resulting $0.20$B model retaining $90\%$
of baseline QPS.

Vocabulary scaling with PCGM provides the largest increase in
revenue lift within the scaling group. It expands the model from
$0.20$B to $0.46$B and raises the cumulative improvement to
$+0.85\%$, while QPS recovers from $90\%$ to $95\%$.
PCGM replaces the dense FFN gate projection with prefix-indexed
memory lookup, expanding parametric memory while reducing part
of the dense computation. The observed throughput improvement
is consistent with this computation--memory trade-off.

Stream scaling with diagonal Hyper-Connections further improves
revenue by expanding residual information pathways while keeping
each attention and FFN sublayer evaluated only once. The resulting
$0.49$B model achieves a $+1.13\%$ cumulative improvement at
$90\%$ of baseline QPS. Decision-depth scaling with SARP then
raises the improvement to $+1.41\%$ at $85\%$ QPS, with no visible
parameter-count change at the reported precision. SARP also
reduces the first-level SID loss by approximately $3\%$,
consistent with improved first-level prediction.

% These results suggest that parameter count alone does not fully
% characterize the effectiveness-efficiency trade-off. Input and
% decision-depth scaling improve revenue with little change in
% parameter count, while PCGM substantially expands memory capacity
% without a proportional reduction in throughput. These findings
% favor complementary capacity scaling over dense backbone expansion,
% with each sequential gain measured conditional on the preceding
% variants rather than in isolation.

These results suggest that parameter count alone does not fully capture
the effectiveness-efficiency trade-off. Input and decision-depth scaling
improve revenue with minimal parameter changes, while PCGM substantially
expands memory capacity without proportionally reducing throughput.
The findings favor complementary capacity scaling over dense backbone
expansion; each gain is measured conditional on prior
variants, not in isolation.

\subsection{Evaluation of Value Alignment}
\label{sec:exp:alignment}
We measure alignment using HitRate@\(K\) (HR@\(K\)). For requests sampled
from online logs, we use the complete candidate information at the final
ranking stage, including eCPM and generation scores at each SID level.
HR@\(K\) is the overlap between the top-\(K\) advertisements by eCPM and
by generation score, divided by \(K\), averaged over requests. We evaluate
\textbf{Lvl-1}, \textbf{Lvl-1\&2}, and \textbf{Full}, using the first,
first two, and all \(L\) SID levels, respectively. Base includes all
GR4AD-X optimizations except UniVGD.

\begin{table}[t]
  \centering
  \caption{HitRate@\(K\) against eCPM at three SID scopes. Base uses
  generation probability; UniVGD uses probability at Lvl-1 and the
  RRF beam-selection score thereafter. Full excludes final reranking.}
  \label{tab:alignment}
  \setlength{\tabcolsep}{5pt}
  \renewcommand{\arraystretch}{1.0}
  \begin{tabular}{l ccc ccc}
    \toprule
    & \multicolumn{3}{c}{\textbf{Base Model}}
    & \multicolumn{3}{c}{\textbf{+ UniVGD}} \\
    \cmidrule(lr){2-4}\cmidrule(lr){5-7}
    & Lvl-1 & Lvl-1\&2 & Full
    & Lvl-1 & Lvl-1\&2 & Full \\
    \midrule
    HR@10
      & 0.017 & 0.019 & 0.021
      & \textbf{0.145} & \textbf{0.287} & \textbf{0.298} \\
    HR@50
      & 0.071 & 0.075 & 0.078
      & \textbf{0.194} & \textbf{0.329} & \textbf{0.323} \\
    HR@100
      & 0.216 & 0.224 & 0.221
      & \textbf{0.306} & \textbf{0.441} & \textbf{0.427} \\
    \bottomrule
  \end{tabular}
  \vspace{-16pt}
\end{table}

\noindent\textbf{Baseline misalignment.}
The left columns of Table~\ref{tab:alignment} show that the base model has
low alignment across all three scopes. HR@10 is \(0.021\), indicating that only approximately
\(2.1\%\) of the top-10 eCPM advertisements are also ranked within the top
10 by generation score. More importantly, the three scopes
produce similar results across all evaluated thresholds. This consistency
suggests that the mismatch is not concentrated at a particular SID depth and
cannot be addressed by modifying only one decoding level. Instead, it
reflects a global discrepancy between the next-token prediction objective and
commercial value, motivating value-alignment interventions throughout the
generation process.

\noindent\textbf{Alignment after UniVGD.}
The right columns of Table~\ref{tab:alignment} report the results after
applying UniVGD. UniVGD combines generation probabilities with discriminative scores during generation to obtain the final generation score. Full HR@10 increases from \(0.021\) to \(0.298\), an
approximately \(14\times\) relative improvement. Even under the most conservative setting,
Lvl-1 HR@100 increases from \(0.216\) to \(0.306\).

The Lvl-1 results further distinguish the effects of the pre-generation and
in-generation components. PV-BeamSearch deliberately does not apply
discriminative fusion at the SID level \(1\), as described in
Section~\ref{sec:univgd:in}. Nevertheless, Lvl-1 HR@10 increases from
\(0.017\) to \(0.145\). This improvement originates from PV-RSPO, which
propagates prefix-level eCPM rewards through the SID tree. The generative distribution at each layer is recalibrated by modulating the shared user representations and the SID embeddings used as decoder queries. These results demonstrate that the
two mechanisms are separable and complementary: PV-RSPO improves the model's
internal value representation, while PV-BeamSearch additionally rescues
high-value prefixes that remain underestimated by generation probability
during decoding.

UniVGD reuses the decoder backbone with essentially unchanged parameter
count and one additional token-generation step for reranking. It retains
\(75\%\) of Base QPS; cumulative revenue lift over GR4AD rises from
\(2.00\%\) to \(3.07\%\) (Table~\ref{tab:main-ablation}), a difference of
\(1.07\) percentage points.

\section{Conclusion}
\label{sec:conclusion}
We presented GR4AD-X, a production generative advertising recommendation system that advances reliable SID construction, serving-efficient capacity scaling, and value-aligned generation. Multi-Path SID improves representation reliability by retaining
alternative first-level assignments and completing each with a separate residual quantization trajectory. Multi-dimensional capacity scaling strengthens model capability under strict serving constraints, while UniVGD aligns generation with commercial value through prefix-level reward optimization, value-guided beam pruning, and discriminative reranking. Deployed in Kuaishou's large-scale advertising system, GR4AD-X supports high-traffic real-time candidate generation. Offline evaluations and large-scale online A/B tests validate each component, with the full system achieving a cumulative revenue improvement of $3.07\%$ over the deployed GR4AD baseline.

\bibliographystyle{ACM-Reference-Format}
\bibliography{references}

\appendix

\section{Experimental Setup}
\label{sec:exp-setup}

\paragraph{Deployment setting.}
All online experiments are conducted on the Kuaishou advertising platform, which serves more than 400 million users. The production system integrates reward estimation, online learning, and real-time indexing for continuous model evolution. Serving optimizations enable more than 500 QPS per NVIDIA L20 GPU with end-to-end latency below 100 ms. The GR4AD baseline refers to the deployed production system~\cite{xue2026gr4ad}.

\paragraph{Online A/B protocol.}
Each variant receives a randomly assigned traffic share of $5\%$ and is evaluated for seven days. All reported revenue improvements are statistically significant at the $95\%$ confidence level, corresponding to $p<0.05$ under a two-tailed $t$-test on daily metric values.

\paragraph{Model configuration.}
The base backbone contains \(0.16\)B parameters. Each SID consists of
\(L=3\) levels, and the first-level codebook contains
\(K_1=16{,}384\) centroids. Multi-path assignment retains \(m=2\)
first-level candidates. Beam search uses width \(k=512\), while
PV-BeamSearch considers a relaxed pool of size \(2k\). The RRF smoothing
constant is \(\kappa=k/4\). SARP uses $K'=256$ coarse prototypes obtained by reclustering
the frozen first-level SID centroids, with temperature
$\tau=1.0$ and cosine-loss weight
$\lambda_{\mathrm{cos}}=0.1$. The multi-stream backbone contains \(n=4\)
residual streams.

% \appendix









\end{document}